\documentclass[12pt,a4paper]{article}

\usepackage[british]{babel}

\usepackage[a4paper,top=2cm,bottom=2cm,left=2.5cm,right=2.5cm,marginparwidth=1.75cm]{geometry}

\usepackage[style=apa, backend=biber]{biblatex} % APA 7th edition style citations using biblatex
\DeclareLanguageMapping{british}{british-apa} % Set language mapping
\DeclareFieldFormat[article]{volume}{\apanum{#1}} % Format volume number

\usepackage{amsmath}
\usepackage{graphicx}
\usepackage[colorlinks=true, allcolors=blue]{hyperref}
\usepackage{hyperref}
\usepackage[title]{appendix}
\usepackage{mathrsfs}
\usepackage{amsfonts}
\usepackage{booktabs} % For \toprule, \midrule, \botrule
\usepackage{caption}  % For \caption
\usepackage{threeparttable} % For table footnotes
\usepackage{algorithm}
\usepackage{algorithmicx}
\usepackage{algpseudocode}
\usepackage{listings}
\usepackage{enumitem}
\usepackage{chngcntr}
\usepackage{booktabs}
\usepackage{lipsum}
\usepackage{subcaption}
\usepackage{authblk}
\usepackage[T1]{fontenc}    % Font encoding
\usepackage{csquotes}       % Include csquotes
\usepackage{diagbox}
\usepackage{float}

\usepackage{setspace}
\usepackage{titlesec}
\titleformat{\section} % Redefine section numbering format
  {\normalfont\Large\bfseries}{\thesection.}{1em}{}

\usepackage{lineno} % Add the lineno package

\rightlinenumbers % Right-align line numbers

\usepackage{float}   % for customizing caption position
\usepackage{caption} % for customizing caption format


\title{Autonomous Money Supply Strategy Utilizing Control Theory}

\author[1]{Yuval Boneh}
\affil[1]{\small Conclave}
\affil[1]{Yuval Boneh: \texttt{yuvi@conclave.io}}

\date{}  % Remove date

\begin{document}
\maketitle

\begin{abstract}
  Decentralized Finance (DeFi) has reshaped the possibilities of reserve banking in the form of the Collateralized Debt Position (CDP). Key to the safety of CDPs is the money supply architecture that enables issued debt to maintain its value. In traditional markets, and with respect to the United States Dollar system, interest rates are set by the Federal Reserve in an attempt to influence the effects of excessive inflation. DeFi enables a more transparent approach that typically relies on interest rates or other debt recovery mechanisms being directly informed by asset price. This research investigates contemporary DeFi money supply and debt management strategies and their limitations. Furthermore, this paper introduces a time-weighted approach to interest rate management that implements a Proportional-Integral-Derivative control system to constantly adapt to market activities and protect the value of issued currency, while addressing observed limitations.
\end{abstract}


%-------------------------------------------
% Paper Body
%-------------------------------------------
\section{Introduction}

Blockchain technology has empowered the development of Decentralized Finance (DeFi), employing programmable smart contracts to enable secure and transparent financial transactions. DeFi has further enabled a proliferation of experimental stablecoin architecutres. Stablecoins are cryptocurrencies designed to trade at par with a reference asset, typically the United States Dollar (USD). Stablecoins are generally defined by a balance of three key values: Security, Scalability, and Decentralization. While the nomenclature may vary, the first principle considerations remain the same, and the general argument is that a stablecoin design can only optimize for two of the three. Compromising decentralization and security introduce existential risk into a protocol, as exhibited by the collapse of Terra's UST stablecoin (\cite{luna-crash}), and BUSD's rapid supply fall attributed to regulatory actions (\cite{busd}). Meanwhile, scalability may not be core to a design, but can be achieved over time as the model develops trust and robustness. To date, the most successful secure and decentralized architecture concept is that of a Collateralized Debt Position (CDP) protocol.

CDPs refer to positions created by locking collateral in a smart contract to generate some kind of debt, typically stablecoins. Stablecoins represent the system's debt, so safeguarding their value is cruitial. Established oracle infrastructures exist to ensure that the value of collateral and debt is known at all times, however the mechanism by which minted debt assets maintain peg varies across designs. Liquity's LUSD stablecoin offers the highest level of decentralization and security, with rigorous mechanisms designed to ensure the value of the LUSD token remains pegged. MakerDAO's DAI stablecoin and Aave's GHO stablecoin have interest rates set by governance bodies, which statically adjust interest rates depending on the need for the stablecoin supplies to contract or expand. This research provides an analysis of existing CDP protocols and respective money supply theories and examines potential shortfalls before proposing a novel design that mitigates identified risks.

\section{Industry Review}

The purpose of this section is to provide an overview of some of the largest decentralized CDP protocols in DeFi and assess their money supply strategies for efficiency. Some mechanisms will be discussed in terms of control theory to provide a standardized method of communicating assessed drawbacks and limitations.

\subsection{MakerDAO}

The Maker Protocol, or Multi-Collateral Dai (MCD) system (\cite{makerWhitePaper}), accepts governance-approved collateral assets and allows users to mint the DAI stablecoin. Opening a CDP with Maker incurs an accruing stability fee, paid in DAI, to the Maker Buffer. In turn, the Maker Buffer funds the DAI Savings Rate (DSR) vault; a vault where users can deposit DAI to earn interest. DAI maintains its price stability by Maker governance adjusting stability fees, as well as the DSR rate, in order to balance supply and demand mechanics. An additional Peg Stability Module (PSM) enables 1:1 exchange with a limited amount of USDC, further tightening the peg by creating rapid arbitrage opportunities when price shifts sufficiently to make the arbitrage transaction profitable.

DAI's stability is contingent on interest rates being modified via governance, so while they are adaptive, the system response is inherently slow and inefficient. Maker governance manages the DAI borrow rates of individual collaterals, as well as the DSR rate, resulting in a cumbersome governance liability. An April 2024 governance proposal (\cite{makerGov}) included 10 independent interest rate modifications for the DAI ecosystem based on observations made by BA Labs. There is a significant amount of research and data accompanying each proposal, presumably consuming a large quantity of resources, akin to the US Federal Reserve's approach to interest rate management. Maker sets the baseline for CDP technology, deviating from the traditional gold-backed financial system only insofar as transparency and the use of secure and decentralized blockchain technology.

\begin{figure}[!ht]
  \centering
  \includegraphics[width=1\linewidth]{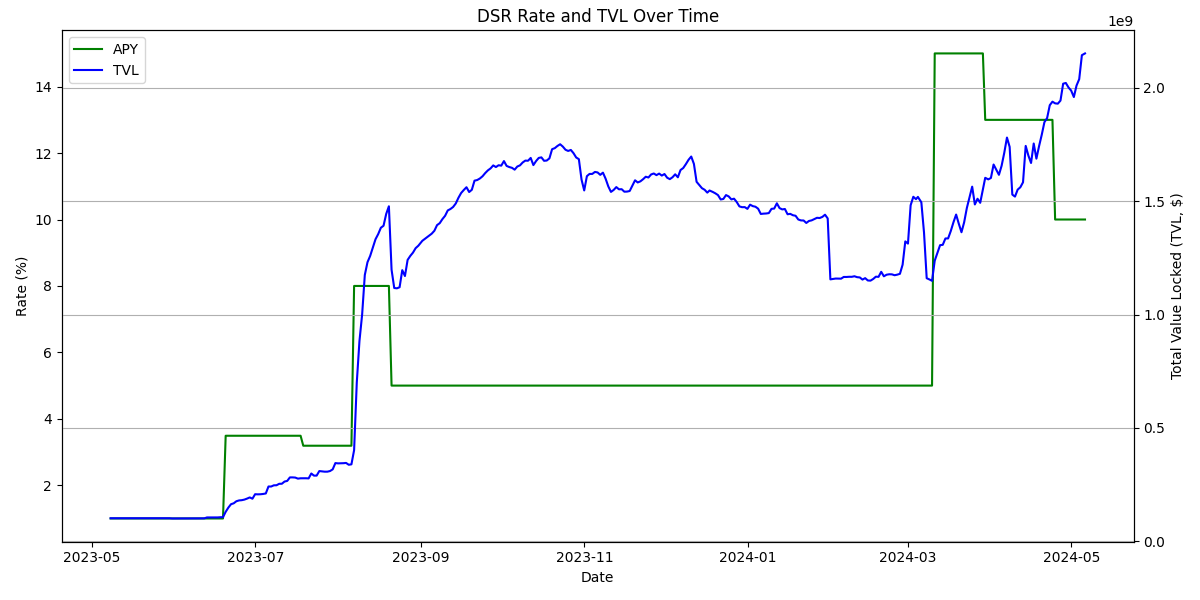}
  \caption{\label{fig:dai-dsr}Maker's DAI Savings Rate historical interest rates.}
\end{figure}

In August 2023, Maker drastically increased their DSR savings rate\footnote{Source: \url{https://defillama.com/}} (Figure \ref{fig:dai-dsr}), effectively prodiving a conduit between their large off-chain reserves earning significant interest in a high rates environment, and the on-chain intrest rate environment.

This transition rippled throughout the CDP landscape and the instability it introduced provides an excellent case study for building a more resilient system.

\subsection{Liquity V1}

The Liquity V1 Whitepaper (\cite{liquityV1}) explains their approach to protecting the peg and the value of their decentralized stablecoin LUSD by way of liquidation and redemption mechanisms. Of note is the redemption mechanism, which sets a floor price for LUSD should the price fall below \$1. The system allows LUSD holders to redeem their LUSD for underlying ETH collateral based on the face value of the redeemed tokens, the current ETH:USD rate, and the current base rate. The system uses the LUSD to repay debt of the riskiest position (lowest collateral ratio), and transfers the respective amount of ETH to the redeemer. The process is fair (the value of ETH transferred from the position is equal to the amount of LUSD token debt paid) and borrowers do not suffer a net loss from being redeemed. Redemptions are subject to a redemption fee, which is a function of the base rate and the redeemed amount of LUSD, with a minimum redemption fee of 0.5\%. Therefore, redemption only becomes profitable when LUSD falls below \$0.995, protecting borrowers from constant redemptions when LUSD value is maintained at \$1.

\begin{figure}[!ht]
  \centering
  \includegraphics[width=1\linewidth]{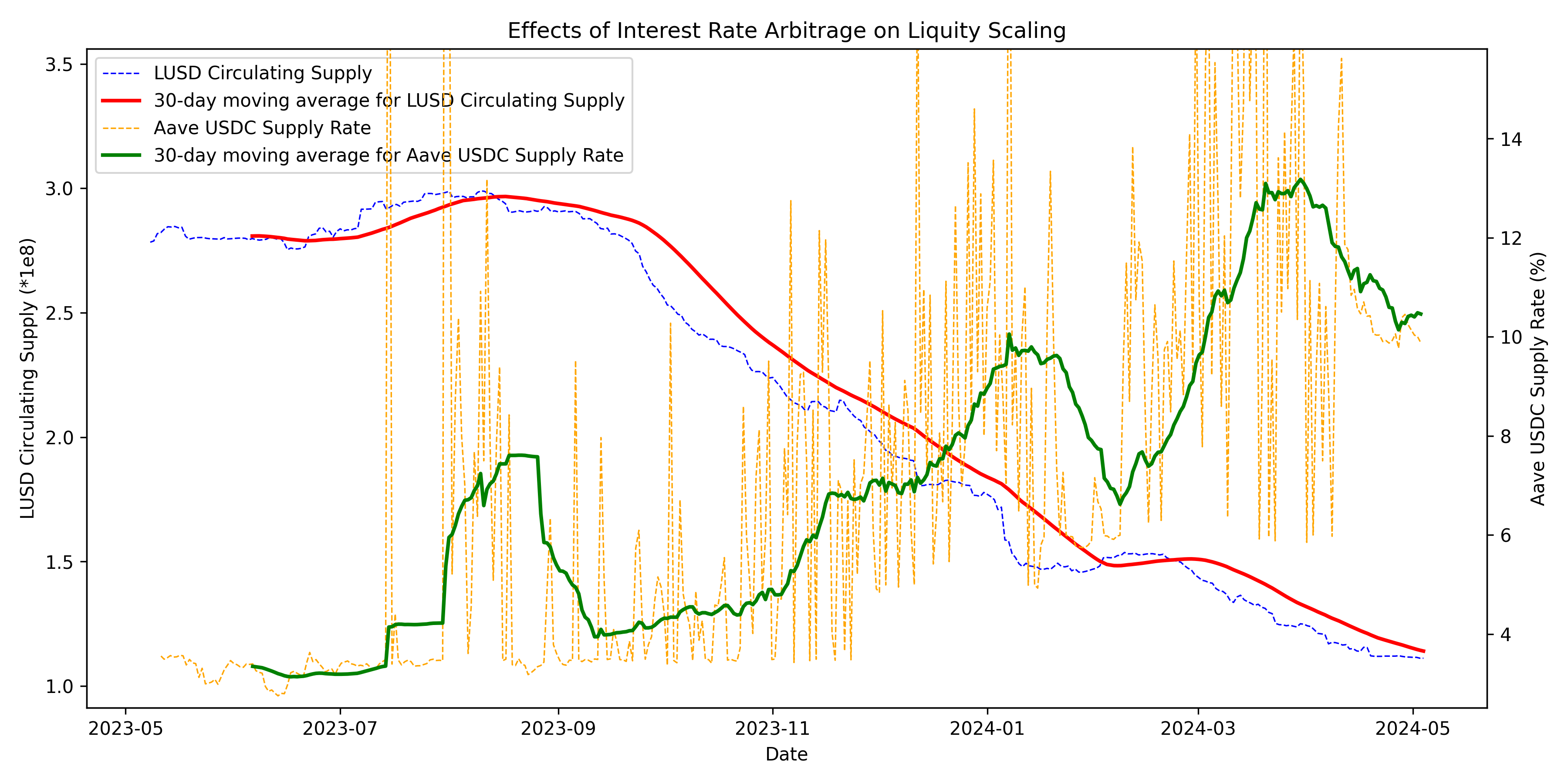}
  \caption{\label{fig:lusdVsRates}Effects of Interest Rate Arbitrage on Liquity Scaling.}
\end{figure}

Liquity offers zero-interest lending, so as broader interest rates rose, the cost of interest rate arbitrage remained the same while profitability increased. As a result, LUSD supply shrank dramatically over the last 12 months, for reasons that will be explained. Aave USDC supply on the Ethereum Mainnet is typically considered the most accessible, lowest risk, stablecoin strategy, and is thus compared as the risk-adjusted baseline rate for DeFi. Data can be observed in Figure \ref{fig:lusdVsRates}, which demonstrates LUSD circulating supply\footnote{Source: \url{https://defillama.com/}} and Aave USDC supply rates\footnote{Source: \url{https://aavescan.com/}} over the last 12 months. Strategic users deposited large amounts of collateral into Liquity to mint LUSD and sell it to the market in favour of stablecoins that could earn high interest elsewhere. This behaviour generated sustained selling pressure on LUSD, which led to consistently profitable redemptions. Consequently, in order to borrow LUSD, users required increasingly high collateral ratios, which inhibits the capital efficiency of the protocol. The interest rate arbitrage became inefficient and users exitted Liquity V1, shrinking its supply.

To be clear, Liquity V1 has always worked exactly as intended. Liquity V1 set a new standard in decentralized CDP architecture, and was subsequently forked by many teams. While the Liquity V1 money supply model works exceptionally well in a low interest rate environment, it can not adapt to a high interest rate environment. Developing a system that is resilient to to the arbitrage opportunities around it requires an adaptive cost of borrowing, or interest rate.

\subsection{Aave GHO}

GHO is a decentralized and collateralized stablecoin native to the Aave protocol (\cite{aaveGho}). GHO has similar supply functionality to Maker in that users must deposit collateral in order to mint the stablecoin, however the GHO money market integrated into Aave performs a passively scalable DSR role. Since the potential supply of GHO is unlimited (ignoring their programmable cap), conventional lend/borrow interest rate models that reference utilization do not apply to the CDP borrow rate. Instead, the GHO interest rate is set by Aave Governance, which "statically adjusts interest rates depending on the need for the GHO supply to contract or expand" (\cite{aaveGhoIR}). As such, Aave's stablecoin has the ability to respond to changes in market dynamics by way of governance voting for adjustments to the global interest rate. This overcomes Liquity V1's inability to restore supply dynamics, but introduces other limitations.

There is a distinct relationship between the cost of borrowing and the price of the stablecoin. This is a known product of interest rate arbitrage. When interest rates increased broadly (exemplified on-chain by the DSR increase), opportunities for profit presented themselves, and the low-interest \$GHO stablecoin faced the same sell pressure as \$LUSD. Since Aave employs a global interest rate in lieu of a redemption mechanism, the \$GHO stablecoin lost peg, falling to a price as low as approximately \$0.95. Forum posts as far back as August of 2023 propose increases in the interest rate in response to market activities (\cite{aaveForumIR}), with eight additional proposals following in as many months (Table \ref{GHOtable}).

\begin{table}[!ht]
  \caption{\label{GHOtable}Sample history of GHO borrow rate adjustments}
  \begin{threeparttable}
    \begin{tabular*}{\columnwidth}{@{\extracolsep\fill}llll@{\extracolsep\fill}}
      \toprule
      Date & Adjustment & Forum Post\\
      \midrule
      04 Apr 24 & from 7.92\% to 10\%\tnote{1} & GHO Stewards - Increase GHO Borrow Rate to 10\%\\
      13 Mar 24 & from 7.92\% to 13\% & GHO Stewards + Borrow Rate Update\\
      08 Mar 24 & from 7.48\% to 7.92\% & Increase GHO Borrow Rate 08-03-2024\\
      29 Feb 24 & from 6.35\% to 7.48\% & Increase GHO Borrow Rate 29-02-2024\\
      04 Dec 23 & from 5.22\% to 6.35\% & Increase GHO Borrow Rate 100 bps to 6.35\%...\\
      21 Nov 23 & from 4.72\% to 5.22\% & Increase GHO Borrow Rate to 5.22\% on Aave V3\\
      30 Oct 23 & from 3.00\% to 4.72\% & GHO - Increase Borrow Rate\\
      05 Oct 23 & from 2.50\% to 3.00\% & Further increase GHO Borrow Rate\\
      23 Aug 23 & from 1.50\% to 2.50\% & Increase GHO Borrow Rate\\
      \bottomrule
    \end{tabular*}
    \begin{tablenotes}
      \item Source: \url{https://governance.aave.com/c/governance/}
      \item[1] This proposal was submitted as a progressive step towards the previous 13\% proposal.
    \end{tablenotes}
  \end{threeparttable}
\end{table}

\begin{figure}[!ht]
  \centering
  \includegraphics[width=1\linewidth]{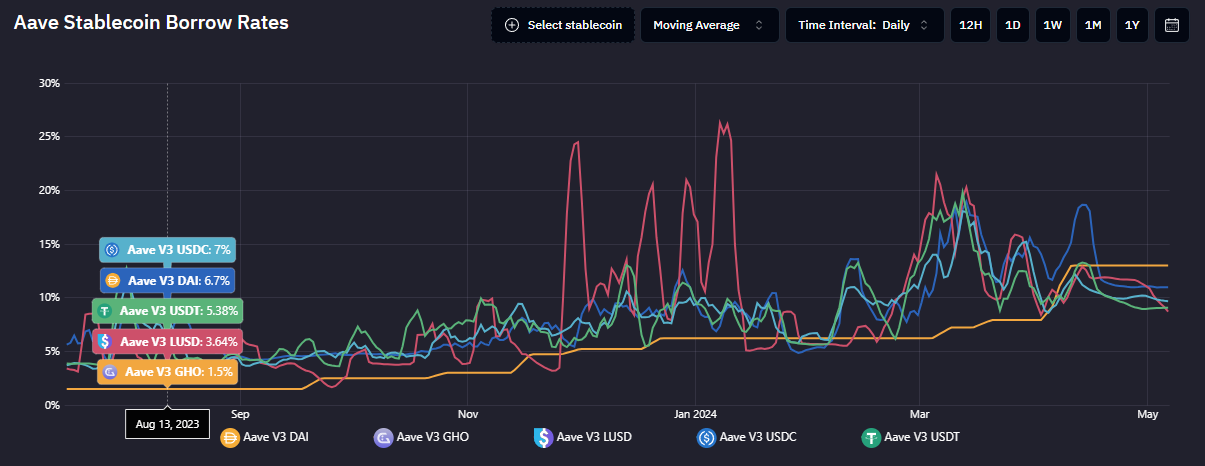}
  \caption{\label{fig:ghoBorrowRates} Historical GHO Borrow Rates (\cite{ghoAnalytics}).}
\end{figure}

Historical GHO borrow rates can be observed and compared to other borrow rates for stablecoins on Aave V3 in Figure \ref{fig:ghoBorrowRates}.

\begin{figure}[!ht]
  \centering
  \includegraphics[width=1\linewidth]{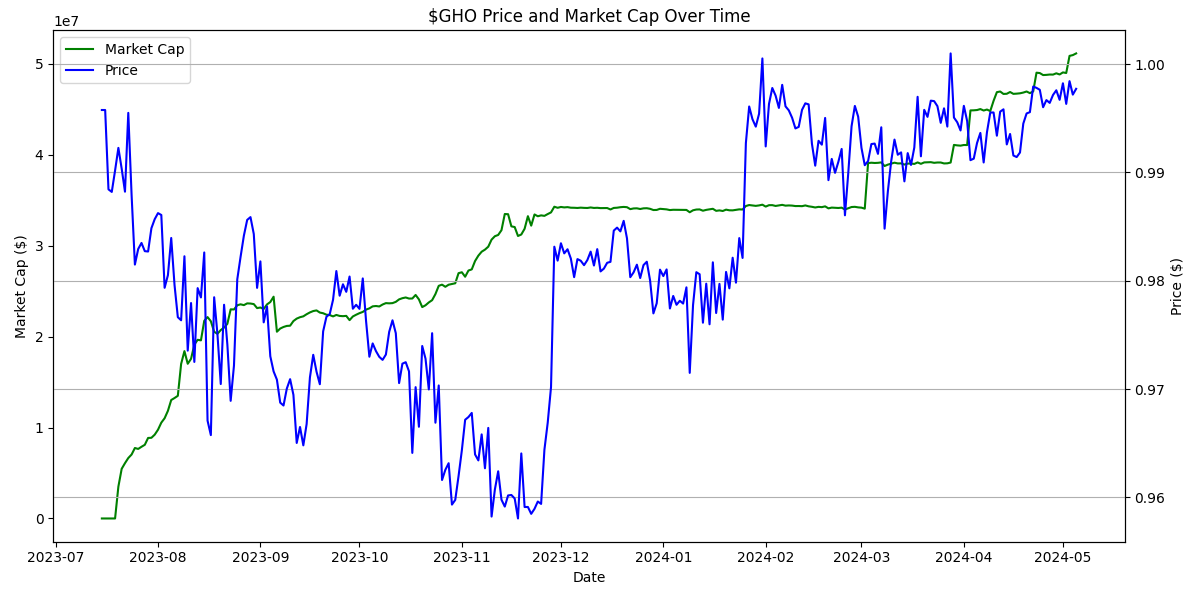}
  \caption{\label{fig:ghoData} Historical GHO Price and Market Cap over time.}
\end{figure}

The delayed nature of Aave governance's response is evident as market rates rise, as well as fall, whereby the interest rate for borrowing GHO is almost always the cheapest until approximately April 2024, after which it is now the most expensive, pending further governance proposals to adjust the interst rate. The effect of such a delayed response is present in GHO's market price, which has consistently been traded off in favour of profitable opportunities, while its market capitalization (and thus circulating supply) has been steadily increasing\footnote{Source: \url{https://www.coingecko.com/}} (Figure \ref{fig:ghoData}).

Going forward, for as long as borrowing GHO remains comparably more expensive, it will face a challenging effort to scale. Adaptive interest rates mitigate the problems faced by Liquity V1, but reliance on governance is slow, cumbersome, and the effective control system response is inhibitingly slow.

\subsection{Liquity V2}

Liquity V2 (\cite{liquityV2WhitePaper}) aims to address the delay by introducing user-set interest rates (\cite{liquityV2}). Liquity V2 redemption mechanics target individual interest rates instead of collateral ratios. This means that if users set their interest rates high enough, they can access the same capital-efficient loans that V1 offered. It can be expected that Liquity interest rates will rise and fall in tandem with the market's, within a range of low arbitrage profitability. However, there remain several drawbacks in the Liquity V2 model.

\begin{figure}[!ht]
  \centering
  \includegraphics[width=0.8\linewidth]{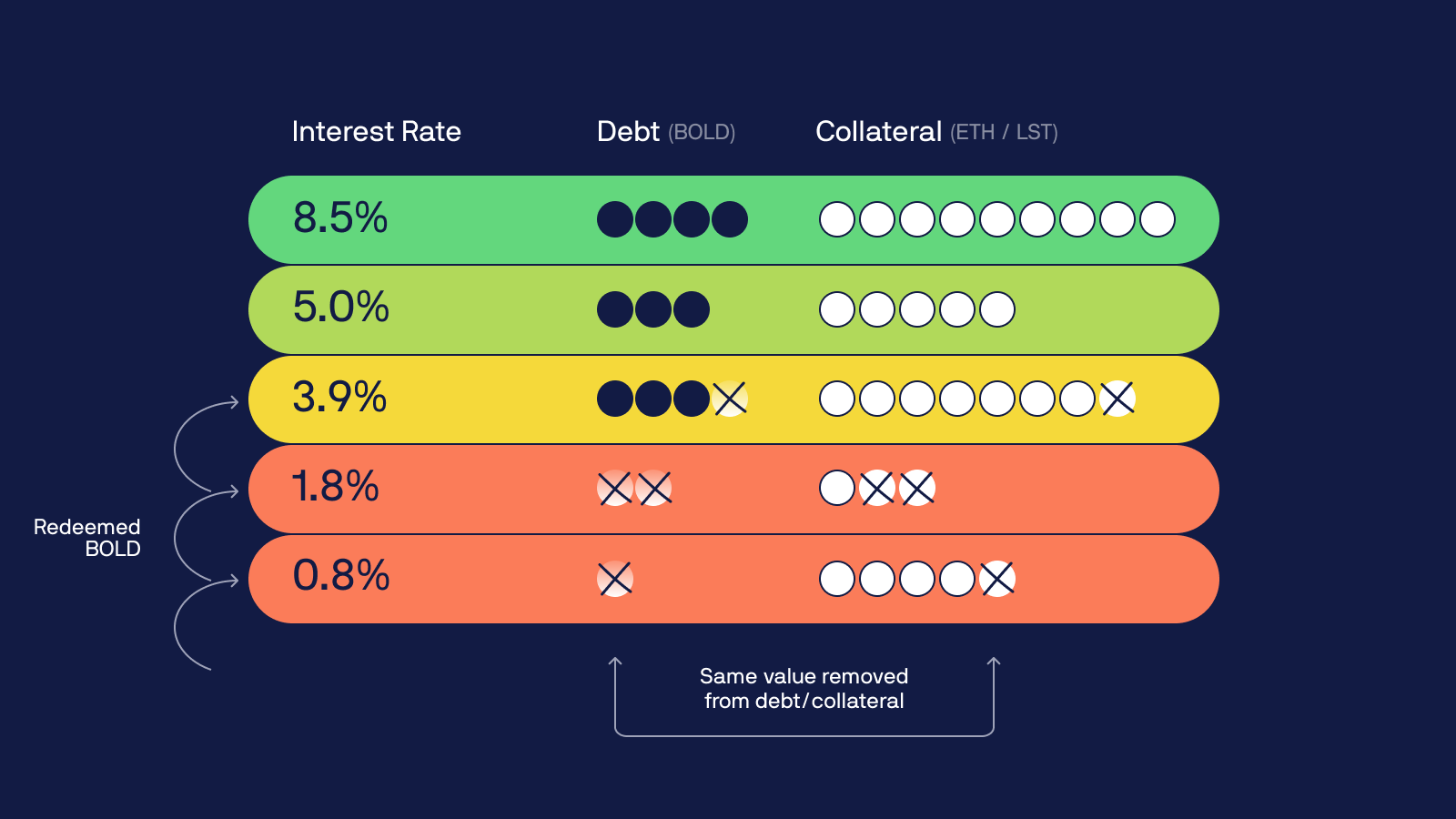}
  \caption{\label{fig:liquityV2Redemption} Liquity V2 redemption scenario (\cite{liquityV2}).}
\end{figure}

The first is that in this design, interest rates do not necessarily mitigate the risk of redemption. \textcite{liquityV2} describes a redemption scenario in Figure \ref{fig:liquityV2Redemption}, in which borrowers are ordered by interest rate with their debt and collateral values expressed as white and black circles. A redemption of four stablecoin units repays the debt of the two lowest interest rate positions and part of the 3.9\% interest rate position, leaving additional collateral behind for position owners.

It is apparent that maintaining an interest rate that is not the lowest does not protect users from redemption, as the size of redemption can not be guaranteed. A smaller redemption would leave the 3.9\% interest rate position unaffected, while an eight unit redemption would affect the 5\% interest rate position as well. The result is that position managers will need to consider existing positions and potential redemption sizes when managing interest rates. While it is reasonable that the base rate should be influenced by the actions of users with highest risk appetite, the redemption mechansim may continue to adversely impact users whose actions remain aligned with the protocol.

\textcite{liquityV2-2} identifies that Liquity V2 infrastructure allows borrowers to delegate their interest rates to third parties, smart contracts, or externally owned accounts. For many users this will provide a solution to the redemption problem by enabling automation of interest rate management to avoid redemption. However, this introduces dependencies into the system, and does not mitigate the requirement to monitor exisitng positions and potential redemption sizes in order to optimize. The alternative is to pay more interest than would otherwise be necessary. Additionally, the efficiency of third party interest rate managers remains to be seen. Regardless, users who want an optimized interest rate will require third party management, which must always consider the possibility of redemption, while users who want a fixed interest rate will have to pay above the base Liquity V2 interest rate.

Finally, the design can be analysed in terms of control theory. When market rates increase, there is a tendency to take on low interest debt in order to profit from interest rate arbitrage. This involves selling the \$BOLD stablecoin for others that earn more interest. As the price of \$BOLD decreases below \$1, redemptions become profitable, prompting Liquity V2 users to increase their interest rates, thus restoring interest rate parity. There exists a relationship between the price of \$BOLD and the perceived 'safe' interest rate, where users are comfortable with their exposure to redemption. The primary inefficiency of the control system lies in the settling time of the market response that corrects the interest rate delta. As rates go up, Liquity V2 will delay in response as users wait until their positions are at threat of redemption before volunteering to pay more interest. The rolling threat of redemption will force an overshoot whereby users increase their interest rates by more than necessarily required in order to avoid having to continually manage their positions. Similarly, on the way down, users will pay more interest than they necessarily need to, while they wait for other positions to buffer their redemption exposure. These effects will not necessarily affect all Liquity V2 users, however they will affect the system's efficiency. The functional tradeoff is that users will always pay more interest than they necessarily need to, so that the system can offer fixed interest rate lending, governed by the profitability of redemption.

In order to improve the efficiency and scalability of the CDP construct, focus should be on a global interest rate solution with minimal response time. Increasing the system's ability to respond to market dynamics will reduce the likelihood and duration of meaningful depeg, eliminating the need for redemptions, and further improving the system's efficiency. Features such as fixed interest rates reduce the core product's ability to optimize for money supply and should be considered later as derivative products.

\subsection{Curve crvUSD}

Curve's crvUSD successfully utilizes an adaptive interterest rate to influence the cost of borrowing. The equation that governs the interest rate for each market is derived in their documentation (\cite{crvUSD}). The interest rate formula scales an exponential, in which the exponent is a function of crvUSD price and a value; sigma. Sigma is a variable that can be configured by the Curve DAO, such that a lower value makes the interest rates increase and decrease faster as crvUSD loses and gains value respectively, and vise versa for higher values.

The strategy is further reinforced by a system called PegKeeper. A PegKeeper is a contract that helps to stabilize crvUSD price by trading crvUSD with counterassets. When the price of crvUSD in a pool is above \$1, PegKeepers may take on debt by minting uncollateralized crvUSD and depositing into specific pools. This increases the balance of crvUSD in the pool, which decreases crvUSD's price. If a PegKeeper has taken on uncollateralized debt and deposited crvUSD into a pool, it may withdraw that crvUSD when the price is below \$1. By withdrawing, the crvUSD token balance will decrease and the price of crvUSD increases. PegKeepers only provide a small buffer because they cannot be allowed to mint crvUSD ad infinitum. This is a simple algorithmic price arbitrage that frontruns the market's response to interest rate changes.

\begin{figure}[!ht]
  \centering
  \includegraphics[width=1\linewidth]{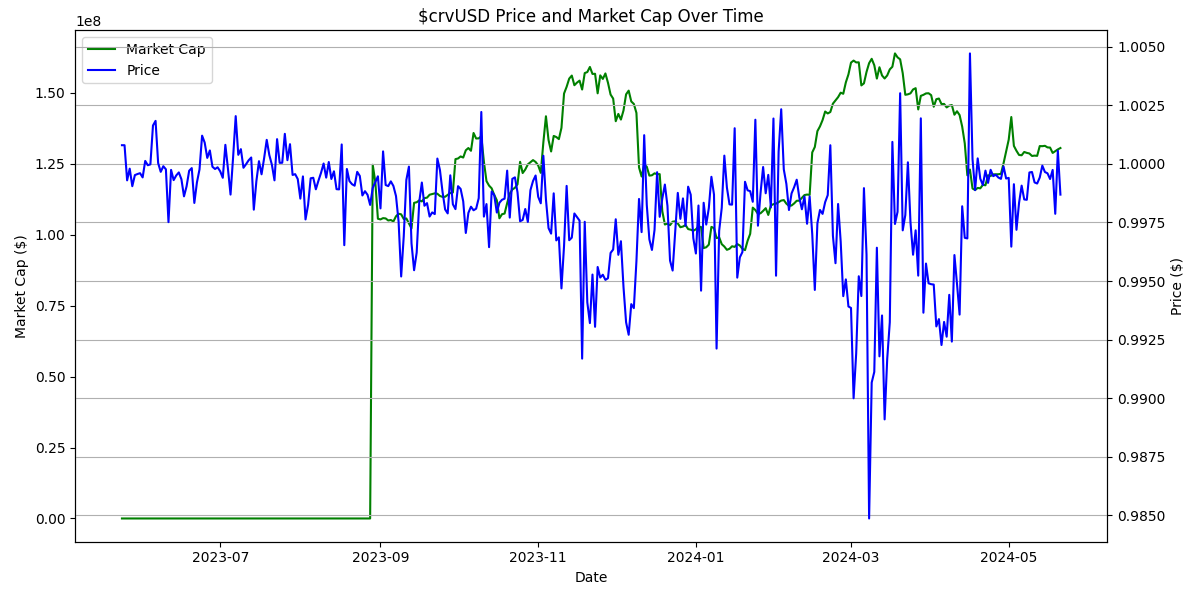}
  \caption{\label{fig:crvUsdData} Historical crvUSD Price and Market Cap over time.}
\end{figure}

Curve's stablecoin system works reasonably well, as can be seen in the data representing the market capitalization and crvUSD price over the last eight months\footnote{Source: \url{https://www.coingecko.com/}} (Figure \ref{fig:crvUsdData}).

Due to limited data, the crvUSD response can only be assessed since August 2023 when MakerDAO's first large rate hike occurred, so the system's response to large systemic changes cannot be assessed because it didn't face the same transition as others. However, its ability to respond to consistent changes since then has been generally promising. The crvUSD stablecoin has remained within a reasonable range of \$1 and the interest rates for borrowing remain competitive.

Overall, the crvUSD design is considered to be the most advanced money supply approach of those examined. That said, the crvUSD money supply strategy still presents interesting limitations. The crvUSD stablecoin relies on debt caps to protect the DebtFraction parameter (a ratio of the PegKeeper's debt to the total outstanding debt) utilized in the interest rate function. As well as somewhat limiting the system's scalability, this creates an artificial utilization signal that informs the interest rate, and is subject to sustained offset. If the PegKeeper has no counterassets with which to purchase debt and the market has an appetite to maintain risk, the crvUSD price can remain offset from the peg. This is mitigated by the sigma variable, however requires governance intervention, which would introduce the same delay as has been observed in Aave's GHO. In its current state, the crvUSD interest rate strategy has no autonomous way of responding to sustained offset. This can be achieved with a Proportional-Integral-Derivative (PID) controller, utilizing control theory.

\section{PID Controller Money Supply Strategy}

\begin{figure}[!ht]
  \centering
  \includegraphics[width=1\linewidth]{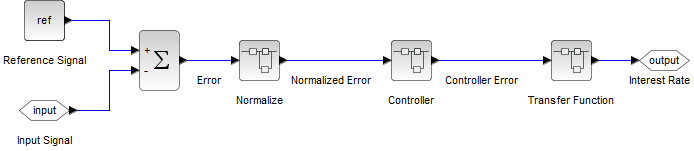}
  \caption{\label{fig:model}General representation of controller logic.}
\end{figure}

This section introduces a control system for money supply management that prioritizes the efficiency of the system's response to market dynamics. The architecture of the controller itself is closely aligned with a previous controller architecture (\cite{mmPid}), modified specifically for this application. Figure \ref{fig:model} provides a general representation of the controller logic.

In general terms, the controller normalizes an error signal to standardize the strategy for general use. It then passes the normalized error through a PID controller that processes independent signals according to Equation \ref{eqControllerError} to output a modified controller error signal.

\begin{equation}
  E_{controller} = E_{P} + E_{I} + E_{D},
  \label{eqControllerError}
\end{equation}

where $E_{controllerError}$ is the modified error (controller output), and $E_{P}$, $E_{I}$, and $E_{D}$ are the proportional, integral, and derivative components of error, respectively.

The controller error then passes through a transfer function, which defines the relationship between the controller error signal and the interest rate.

\subsection{Reference and Input}

The response mechanisms discussed in the previous section are generally dependent on price movements. However, there are many potential leading indicators that may preempt price movements, allowing the system to adjust before market price changes. This is most applicable for StableSwap pools (\cite{stableswap}), in which the invariant allows for extremely low slippage trades even when the pool is imbalanced. For a conventional $xy=k$ constant product pool, imbalances result in impactful slippage, so will reduce the efficiency of the controller. Research by \textcite{detectingDepegs} identifies several strategies that provide leading indicators to price movement. The purpose of their research is to develop off-chain infrastructure to alert liquidity providers of potential depegs, so majority of the leading indicators developed do not suit the purpose of the money supply strategy, however it does attest to the notion of responding to pool imbalances before meaningful price changes. The Gini Coefficient and Shannon's Entropy, described in the paper as a measure of the relative balances of a pool's token, are objective, however are absolute measures of balance. For the purpose of an autonomous on-chain controller input signal and reference, a simple directional asset balance assessment provides a stronger basis.

By assessing the pool balance and targeting a desired weight, a larger weight indicates a surplus of the stablecoin in the pool (excessive supply), and a smaller weight indicates a deficit in the pool (excessive demand). The error signal is defined in Equation \ref{eqError} as the difference between the current weight and the reference (target) weight.

\begin{equation}
  e(w) = w - w_r,
  \label{eqError}
\end{equation}

where $w$ is the weight of the stablecoin in the pool, $w_r$ is the reference (target) weight, and $e(w)$ is the weight error. The controller thus modifies interest rates based on this directional indicator and before the depeg occurs.

Normalizing the ranges $(0, w_r]$ and $(w_r, 1)$ to $(-1, 0]$ and $(0, 1)$, respectively, further standardizes the strategy, making it applicable to any reference weight. The normalized ranges are defined in Function \ref{eqNorm}.

\begin{equation}
  E(w) =
  \begin{cases}
    \dfrac{e(w)}{w_r},       & \text{if } e(w) \leq 0    \\[1em]
    \dfrac{e(w)}{1 - w_r}, & \text{if } e(w) > 0,
  \end{cases}
  \label{eqNorm}
\end{equation}

where $E(w)$ is the normalized error signal with range $(-1, 1)$. This standardization makes the strategy applicable to any reference weight and adaptable for various pool configurations on different exchanges.

\begin{figure}[!ht]
  \centering
  \includegraphics[width=0.7\linewidth]{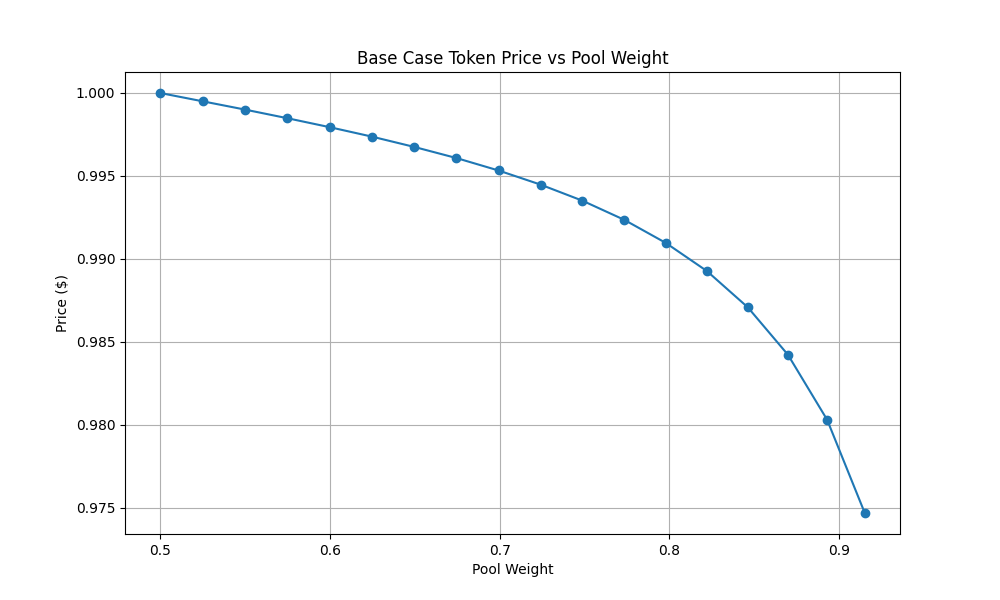}
  \caption{\label{fig:stableswap-price}Example StableSwap price curve, A = 100.}
\end{figure}

A Curve StableSwap pool with amplifiication factor set to 100 has a price curve represented in Figure \ref{fig:stableswap-price}.

With 1,000,000 of token A and 1,000,000 of token B, this pool can facilitate a swap of 400,000 token A for approximately 398,132 token B, resulting in a pool balance of approximately 70\% token A at a price of approximately \$0.995. Commensurate with Liquity V1's redemption profitability case, this scenario forms the base case for the example strategy and tune\footnote{Decentralized Exchanges have varying mechanisms in place for pricing pool imbalances so the degree of acceptable imbalance as it pertains to the base interest rate should be tested thoroughly.}.

\subsection{Transfer Function}

\begin{figure}[!ht]
  \begin{subfigure}{0.49\textwidth}
      \includegraphics[width=\linewidth]{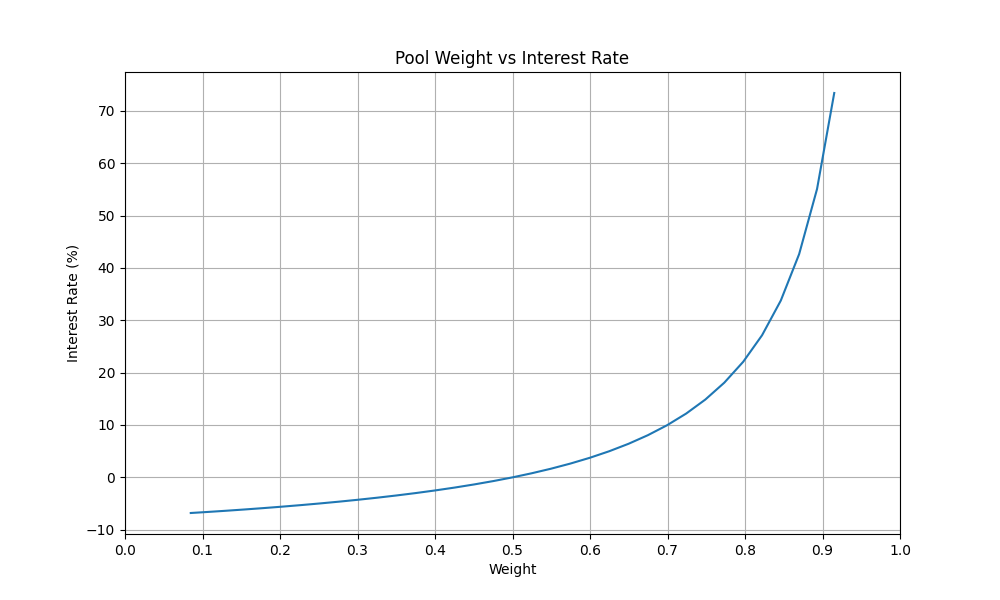}
      \caption{}
  \end{subfigure}
  \hfill
  \begin{subfigure}{0.49\textwidth}
    \includegraphics[width=\linewidth]{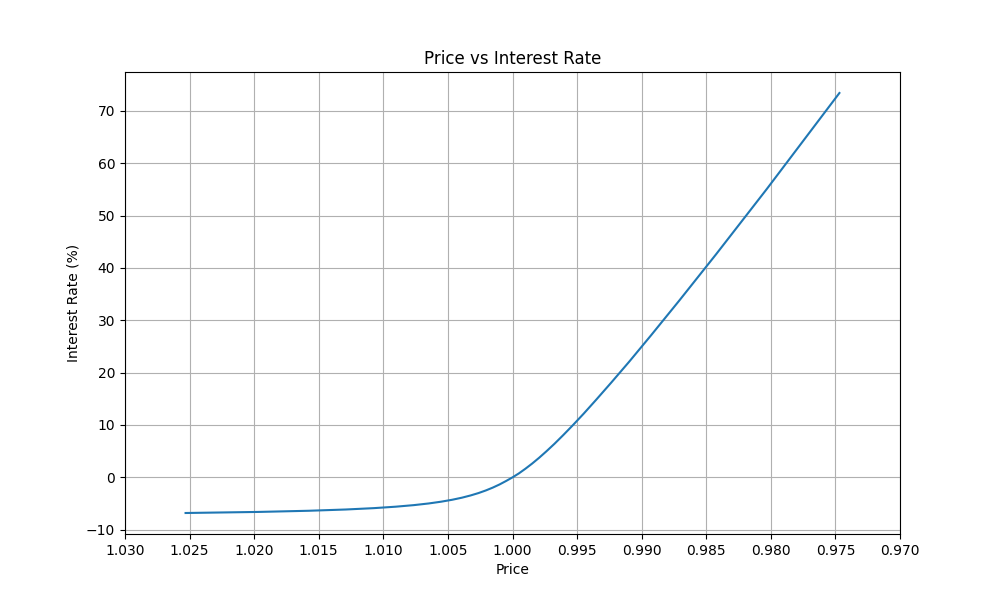}
    \caption{}
  \end{subfigure}
  \caption{Transfer function subset: (a) pool weight vs interest rate, (b) equivalent price vs interest rate.}
  \label{fig:tf_multi}
\end{figure}

The transfer function is derived theoretically on the basis that at 0\% pool imbalance, the output interest rate should be 0\%, and at 100\% pool imbalance (such that the entire liquidity pool has been drained of counterassets, albeit functionally impossible), the output interst rate should be infinite. It is reasonable to allow the system to engage in negative interest rates in order to scale supply for excess demand, however there is no requirement for a negative interest rate to tend towards infinite. As such, a reasonable transfer function to reflect the desired behaviour of the interest rate with respect to pool imbalance is presented in Equation \ref{eqTf}.

\begin{equation}
  r = \alpha \cdot \frac{E_{controller}}{1-E_{controller}},
  \label{eqTf}
\end{equation}

where $r$ is the interest rate, and $\alpha$ is a scaling factor (initially set to $15e-2$). The interest rate curve can be examined in Figure \ref{fig:tf_multi}, noting the plot is a subset of the true pool balance range for demonstration.

The transfer function trends appropriately against pool weight, and the magnitude at 70\% weight is refined to 10\% with $\alpha = 15e-2$. Depending on the volatility and risk profile of the collateral, as well as the pool configuration, $\alpha$ should be varied to set an appropriate base rate.

\subsection{Proportional Component}

In general, the purpose of the proportional component is to tune the magnitude of the initial response, as represented in Equation \ref{eqP}.

\begin{equation}
  E_P = K_P \cdot E(w),
  \label{eqP}
\end{equation}

where $K_P$ is the proportional gain factor. The design of the transfer function is dependent on the range of $E_{controller}$ in that the denominator offsets the signal from the limit in order to define the asymptote. For a proportional gain to be applied, the transfer function would be defined as Equation \ref{eqTfKp}, which would undo the proportional effect of the gain factor and remove the asymptote effect of the curve design.

\begin{equation}
  r = \alpha \cdot \frac{E_{controller}}{K_P-E_{controller}},
  \label{eqTfKp}
\end{equation}

For this reason, $K_P = 1$, and the value of $\alpha$ facilitates a pseudo-proportional component.

\subsection{Integral Component}

The integral component performs a time series analysis of the pool's balance to generate a Time-Weighted Cumulative Error (TWCE) that is scaled by a tunable gain factor (Equation \ref{eqI}).

\begin{equation}
  E_I = K_I \cdot TWCE,
  \label{eqI}
\end{equation}

where $K_{I}$ is the integral gain factor.

$TWCE$ is achieved by multiplying the normalized error signal each time it is reported by the time since it was last reported and adding it to the last TWCE, as summarized in Equation \ref{eqTwce}.

\begin{equation}
  TWCE_i = \sum_{i=1}^{n} E(w)_i \cdot (t_{i} - t_{i-1}),
  \label{eqTwce}
\end{equation}

where $E(w)_i$ is the $i^{th}$ normalized weight error, $t_{i}$ is the time at which the $i^{th}$ error is recorded, and $t_{i-1}$ is the previous time at which error was recorded.

\begin{figure}[!ht]
  \centering
  \includegraphics[width=0.7\linewidth]{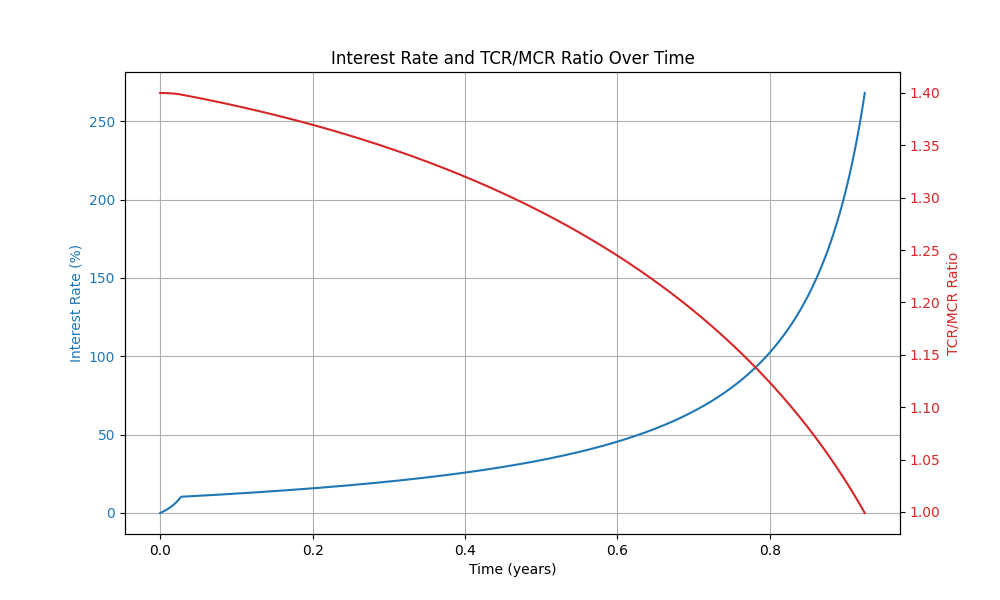}
  \caption{\label{fig:controllerSim140} Interest rate model with $K_I = 1.5$.}
\end{figure}

This rolling calculation is mathematically efficient and can operate effectively ad infinitum. It accumulates when error is positive and dissipates when error is negative, and does so proportionally to the size and sustained time of the error. The effect of cumulative error can be observed in Figure \ref{fig:controllerSim140}, with $K_I = 1.5$, which demonstrates the increase of interest rate leading to complete liquidation in the absence of market action.

Due to the cumulative nature of the integrator component, the modified weight error signal transitions from being sensitive to the proportional component, to being predominantly driven by the integrator component.

\subsubsection{Phi-Strategy}

For a money supply strategy, it is assessed that interest rates should grow or decay at different rates depending on the class of assets in the system. When pool balance tends towards a depeg to the downside, premium, low volatility collateral assets are likely to have a lower Total Collateral Ratio (TCR) relative to Minimum Collateral Ratio (MCR) and should experience slower interest rate growth. Conversely, more volatile assets are likely to have a higher TCR relative to MCR and should experience greater interest rate growth. Therefore, as $TCR/MCR$ increases, so too should $K_I$. While this can be established by tuning $K_I$ for different collateral types, the relationship between $K_I$ and $TCR/MCR$ can be generalized in Equation \ref{eqKi} based on proportionality.

\begin{equation}
  K_I = \phi \cdot (\frac{TCR}{MCR} - 1),
  \label{eqKi}
\end{equation}

where $\phi$ is a scaling factor that governs the aggression of interest rate growth.

Since $TCR/MCR$ evolves over time, Equations \ref{eqI} and \ref{eqTwce} are adapted to form Equation \ref{eqKiTwce} to ensure $K_I$ is updated with each iteration.

\begin{equation}
  E_I = \sum_{i=1}^{n} K_{I,i} \cdot E(w)_i \cdot (t_{i} - t_{i-1}),
  \label{eqKiTwce}
\end{equation}

\begin{figure}[!ht]
  \centering
  \includegraphics[width=0.5\linewidth]{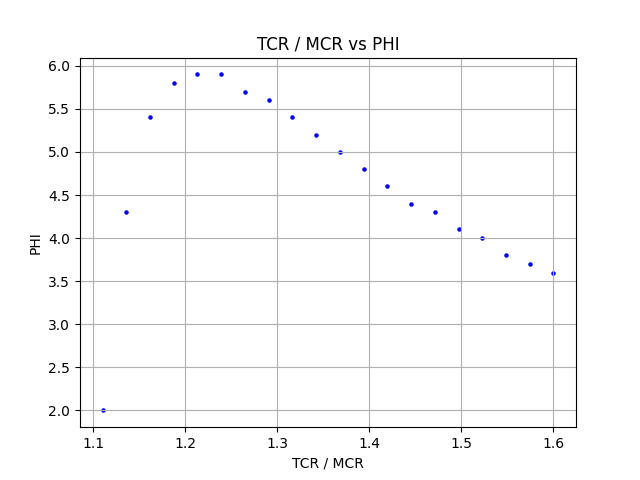}
  \caption{\label{fig:phiSim} Sweep simulation calculating $\phi$ values that return $TCR/MCR$ conditions to $TCR/MCR = 1$ in one year.}
\end{figure}

A dataset can be constructed by sweeping through different collateral ratios and $\phi$ values to identify the $\phi$ value required to return the system to a collateral ratio of one, in one year (Figure \ref{fig:phiSim}). A base rate of 10\% established by the transfer function means that no integrator value is required at $TCR/MCR = 1.1$, so the simulation sweeps from 1.2, and the sensitivity to $\phi$ dissipates after $TCR/MCR = 1.6$.

\begin{figure}[!ht]
  \begin{subfigure}{0.49\textwidth}
      \includegraphics[width=\linewidth]{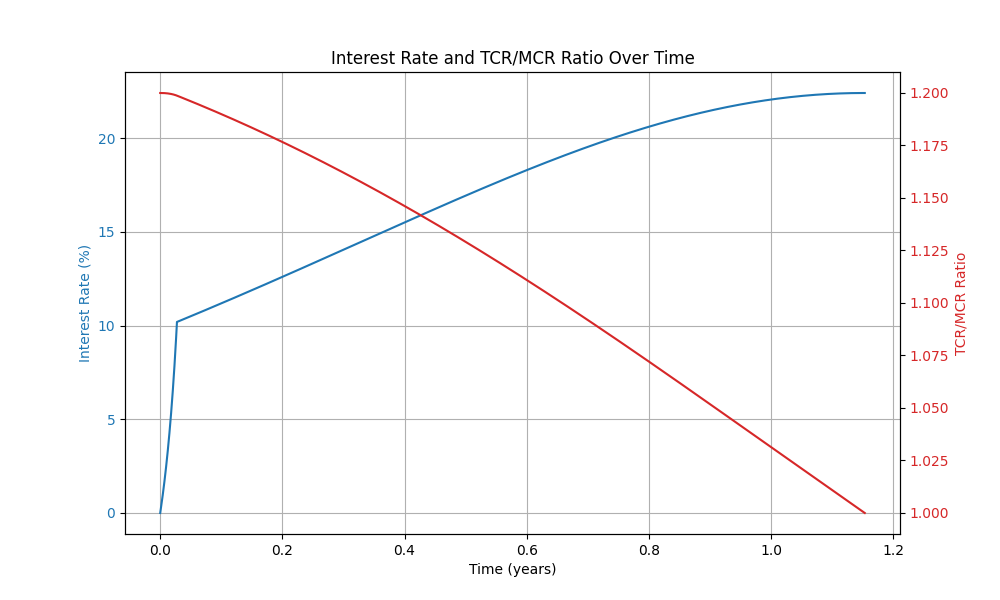}
      \caption{Initial $TCR/MCR$ of 1.2.}
  \end{subfigure}
  \hfill
  \begin{subfigure}{0.49\textwidth}
    \includegraphics[width=\linewidth]{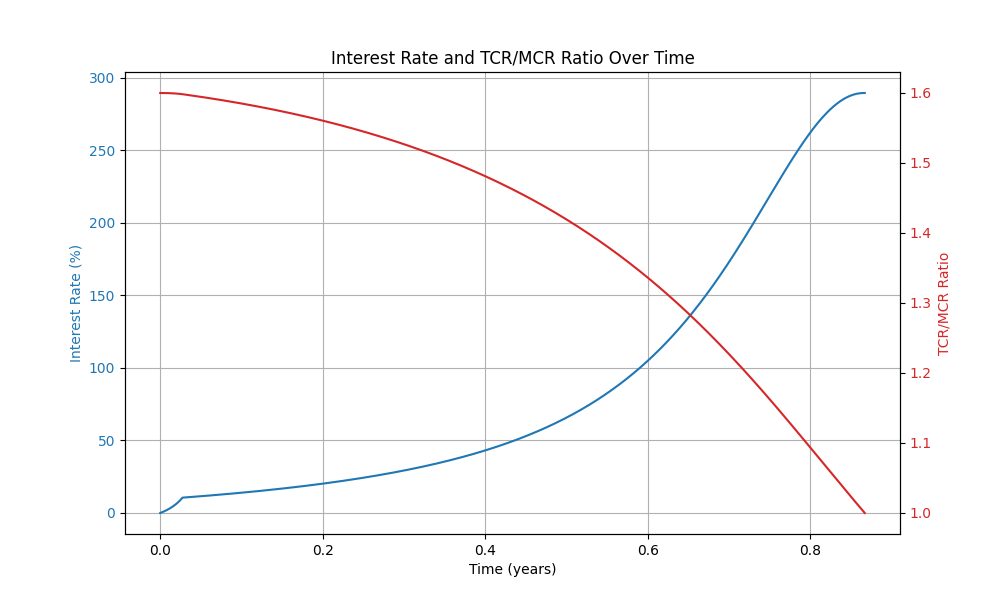}
    \caption{Initial $TCR/MCR$ of 1.6.}
  \end{subfigure}
  \caption{Interest rate controller performance for different $TCR/MCR$ conditions.}
  \label{fig:controllerSim}
\end{figure}

Since any precision of the system's ability to reach a collateralization ratio of one in one year is invalidated by market response, the value of $\phi$ need only be within a reasonable order of magnitude, so $\phi$ is set to $4$. The controller can now be simulated to assess overall effectiveness of returning a CDP to $TCR/MCR = 1$, starting from different points. Simulation outputs are presented in Figure \ref{fig:controllerSim}, in which the pool balance deviates from 0.5 to 0.7 in incremenets of 0.01 and with 12h timesteps. Balance is then maintained at 0.7 for the remainder of the simulation.

In both cases, interest rates initially rise to the same base rate. Once error is maintained at a pool balance of 70\%, the interest rate for the higher $TCR/MCR$ scenario rises significantly faster, accelerating its return to $TCR/MCR = 1$. Since $\phi$ is set to $4$, the two $TCR/MCR$ cases presented recover either faster of slower than the one year target. This is expected from Figure \ref{fig:phiSim}, and ignores market responses, while recovering in a reasonable timeframe. The value of $\phi$ does not necessarily need to be constant, however it is assessed that the added complexity is not justified.

\subsubsection{Mitigating Long-Tail Integrator Risks}

There is potential for the integrator component to accumulate a significant amount of directional bias such that it negates the proportional component and transfer function of the controller. For example if the token weight is under target for a prolonged period of time, the integrator may accumulate significant negative error. Should the pool suddenly shift balances to above target, the proportional component will transition to positive, however the integral component may suppress the summation and result in a reduced overall controller error. This can introduce dangerous attack vectors, so the influence of the integral component must be carefully managed.

This is achieved by programming a limit to the negative accumulation of the integrator component, so that a sudden shift in pool balances can be safely accounted for.

This approach addresses the global rate based on pool imbalance as a priority (pseudo-proportional component), and then increases the burden on riskier assets over time (integral component), ultimately penalizing the least healthy positions first. This eliminates the need for redemptions, so users are free to borrow as they wish and need only focus on their respective position health and interest rate. $K_I$ can be simulated for tuning, or since collateral ratios are inherently normalized, the Phi-Strategy can be deployed to manage different collaterals without modification.

This research presents the case of pool balance shifting to $70\%$ and recovering in isolation in one year, however it should be noted that any further increase in pool balance will compound interest rate growth and result in much faster rate growth. Treating \$0.995 as an example of meaningful depeg, one year is the maximum possible time that it can be sustained without market response. It is expected that as interest rates grow, the market will be influenced to correct the liquidity imbalance, however the controller is designed to correct system leverage in isolation, eliminating any dependency on market activity.

\subsection{Derivative Component}

The derivative component is derived from \textcite{mmPid} and performs a time series analysis of normalized weight error by calculating the gradient over a specified lookback time (Equation \ref{eqD}).

After defining a period, the timestamp of each update can be compared to a delayed TWCE. If at least one period has passed, the delayed TWCE and associated timestamp overwrite the previous TWCE and timestamp, and the current TWCE and timestamp overwrite the delayed TWCE and timestamp. The result is two TWCEs that store the cumulative error one period ago, and now, allowing for the derivative calculation to be completed.

This mitigates the effects of very sudden pool balance changes that would normally be arbitraged, which would subsequently make interest rates excessively sensitive to market actions.

\begin{equation}
  E_{D} = K_{D} \cdot \frac{TWCE_{delayed} - TWCE_{previous}}{t_{delayed} - t_{previous}},
  \label{eqD}
\end{equation}

where $K_{D}$ is the derivative gain coefficient, $TWCE_{delayed}$ is the most recent TWCE recorded before the current value, and is only updated after at least one period, and $TWCE_{previous}$ is the $TWCE$ recorded at least one period before $TWCE_{delayed}$, and allows for the derivative calculation. $t_{delayed}$ and $t_{previous}$ are the respective timestamps.

The derivative component allows the system to compensate for short term (commensurate with the specified lookback time) market actions, to curb rapid deviations from reference weight, protecting the protocol and users.

$K_D$ should be tuned according to the protocol's risk tolerance under different market scenarios, with consideration for arbitrage opportunities and peg support infrastructure.

In essence, the controller modifies the weight error to account for the size of the error, how long the error has been sustained, and how quickly the error was incurred.

\section{Implementation}

This strategy was developed for strict on-chain implementation, with no off-chain dependencies. The response of the controller can be greatly improved with off-chain infrastructure, however for the purpose of this research the trade-off is undesirable.

The arithmetics employed are within the scope of most smart contract programming languages and calculations can be programmed efficiently in order to reduce gas costs.

The pseudo-proportional gain and integral gain may be set algorithmically to achieve generic market responses in the the absense of fine tuning and simulation, or may be set as discrete constants to achieve a finer response for independent markets.

Derivative gain architecture may be comparatively expensive so its use should be considered against the intended network infrastructure. In most cases, it is anticipated that interest rates should not be highly sensitive to derivative gain, and its use is not strictly needed.

Contract architecture has been developed and tested separately and compared in production privately. Some information may be released in future publications.

\section{Conclusion}

The control system described in this paper enables a reserve currency to employ passively adaptive money supply theory with autonomous time-based compensation. There is no requirement for human intervention, and the interest rate updates as regularly as interactions with the protocol. Exact controller configurations have been omitted due to the sensitive nature of the performance of the strategy. As always, security remains the priority, and any contracts derived from this research should be thoroughly and tested and independently audited before implementation.

\printbibliography

\end{document}